\documentstyle[aps,epsf,twocolumn,prl]{revtex}
\def\beq        {\begin{equation}}
\def\eeq        {\end{equation}}
\def\beqn       {\begin{eqnarray}}
\def\eeqn       {\end{eqnarray}}
\def\bmat       {\left[ \begin{array}}
\def\emat       {\end{array} \right]}
\def\barr       {\begin{array}}
\def\earr       {\end{array}}

\def\f		{\frac}

\def\d		{\delta}
\def\e		{\epsilon}
\def\m		{\mu}

\def\s		{\sigma}

\def\bA		{{\bf A}}

\def\bI		{{\bf I}}

\def\bM		{{\bf M}}

\def\bQ		{{\bf Q}}

\def\bT		{{\bf T}}
\def\bU		{{\bf U}}

\def\bg		{{\bf g}}
\def\bk		{{\bf k}}
\def\bl		{{\bf l}}

\def\br		{{\bf r}}

\def\la		{{\langle}}
\def\ra		{{\rangle}}

\def\dr		{\dagger}
\def\up		{\uparrow}
\def\dn		{\downarrow}
\def\cox	{\cos k_x}
\def\coy	{\cos k_y}
\def\sumi	{\sum_i}

\def\sumis	{\sum_{i\s}}

\def\sumijs	{\sum_{\la ij\ra\s}}

\def\sumks	{\sum_{\bk\s}}
\def\cis	{c_{i\s}^{}}
\def\cisd	{c_{i\s}^\dr}
\def\ciu	{c_{i\up}^{}}
\def\ciud	{c_{i\up}^\dr}
\def\cid	{c_{i\dn}^{}}
\def\cidd	{c_{i\dn}^\dr}

\def\cjs	{c_{j\s}^{}}

\def\niu	{n_{i\up}^{}}
\def\nid	{n_{i\dn}^{}}

\def\cks	{c_{\bk\s}^{}}
\def\cksd	{c_{\bk\s}^\dr}

\def\sud	{\f{\s mU}{2}}
\def\ek		{\e_k^{}}
\def\Ekp	{E_k^{\scriptscriptstyle (+)}}
\def\Ekm	{E_k^{\scriptscriptstyle (-)}}
\def\Ekpm	{E_k^{\scriptscriptstyle (\pm)}}

\def\eiky	{e^{ik_y}}
\def\emiky	{e^{-ik_y}}

\def\ni		{n_i^{}}

\newcommand{\av}[1]	{\langle #1 \rangle}

\begin{document}
\draft
\twocolumn[\hsize\textwidth\columnwidth\hsize\csname @twocolumnfalse\endcsname
\title{Antiferromagnetic Correlation under the External Magnetic Field
in the Hubbard Model}
\author{Seung-Pyo Hong, Hyeonjin Doh, and Sung-Ho Suck Salk}
\address{Department of Physics, Pohang University of Science and Technology,
Pohang 790-784, Korea}
\maketitle
\begin{abstract}
By deriving a generalized Harper's equation,
we investigate the effects of temperature $T$ and hole doping $\d$
on antiferromagnetically correlated electrons
under the influence of an applied magnetic field.
We obtain a phase diagram in
a $T$-$\d$ plane for staggered magnetization 
under the external field.
We find that away from half filling 
the maximum values of staggered magnetization occur 
with rapid convergence to a finite temperature
constant near half filling.
We discover the reentrant behavior of the staggered magnetization 
in the presence of the magnetic field.
\end{abstract}
]

Hubbard model is one of the most widely used model for studying 
the systems of correlated electrons.
The original Harper's equation 
\cite{hofstadter,macdonald,hasegawa}.
does not contain the effects of electron correlations 
and thus it takes into account only non-interacting electrons 
under external fields.
To the best of our knowledge,
there is no study of Harper's equation
which takes into account such correlation effects.
By considering an infinite square lattice,
we derive a generalized Harper's equation 
which allows to investigate 
antiferromagnetic electron correlations
under the applied magnetic field.
Earlier, our study was limited only to zero temperature and 
half filling \cite{doh}.
Here we study how 
the applied magnetic field affects the system of antiferromagnetically
correlated electrons
at finite temperatures and away from half filling.

The Hubbard model describing the two-dimensional system (square lattice)
of electron correlations under an external magnetic field
is given by 
\beqn 
H &=& -t\sumijs \exp\left(-i\f{2\pi}{\phi_0}\int_j^i \bA\cdot d\bl\right)
  \cisd\cjs \nonumber \\
  && + U\sumi\niu\nid - \mu\sumis\cisd\cis~~,
\eeqn
where $t$ is the hopping integral; $\bA$, the electromagnetic vector potential;
$\phi_0=\f{hc}{e}$, the elementary flux quantum; 
$U$, the on-site Coulomb repulsion energy,
and $\mu$, the chemical potential.
$\la ij\ra$ stands for summation over nearest neighbor sites $i$ and $j$.
$\cisd$ ($\cis$) is the creation (annihilation) operator of an electron of
spin $\s$ at site $i$, and $\niu$ ($\nid$), the number operator of
an electron of spin up (down) at site $i$.

We are interested in a uniform staggered magnetization 
(antiferromagnetic (AF) order),
\beqn
m =  e^{i\bQ\cdot\br_i}\av{\ciud\ciu-\cidd\cid}~~,
\eeqn
where $\bQ=(\pi,\pi)$,
and $\br_i=(i_x,i_y)$ with $i_x$ and $i_y$ being integers including $0$.
The lattice spacing is set to be a unity.
Introducing the mean field (Hartree-Fock) approximation,
\beqn
\niu\nid 
&\simeq& \av{\ciud\ciu}\cidd\cid + \av{\cidd\cid}\ciud\ciu
  - \av{\ciud\ciu}\av{\cidd\cid} \nonumber \\
&& \hspace*{-1cm} -\av{\ciud\cid}\cidd\ciu - \av{\cidd\ciu}\ciud\cid
  + \av{\cidd\ciu}\av{\ciud\cid}~~,
\label{hf}
\eeqn
and using the Landau gauge $\bA=B(0,x,0)$,
we obtain, in the momentum space,
\beqn
&& H = \nonumber \\
&& -t\sumks\left[2\cos k_x\cksd\cks 
  + e^{-ik_y} c_{\bk-\bg,\s}^\dr\cks 
  + e^{ik_y} c_{\bk+\bg,\s}^\dr\cks\right] \nonumber \\
&& -\f{mU}{2}\sumks \s c_{\bk+\bQ,\s}^\dr\cks 
  + [\f{U}{2}(1-\d)-\mu]\sumks \cksd\cks~~, 
\label{hammag}
\eeqn
where
$\bg\equiv \left(2\pi\f{\phi}{\phi_0},0\right) = \left(2\pi\f{p}{q},0\right)$
with $\f{p}{q}$, the number of flux quanta per plaquette.
For simplicity we considered a uniform doping $\d=1-\av{\ni}$
in the expression above.
The first bracketed term in (\ref{hammag}) represents hopping processes;
the first term in the bracket 
is contributed from the nearest neighbor hopping in the $x$-direction
and the last two terms in the bracket,
from the nearest neighbor hopping in the $y$-direction.
The external vector potential $\bA$ shifts the wave vector of electron 
by $\bg$ during its hopping in the $y$-direction.
The second term results from
the antiferromagnetic electron correlations,
and the third term shifts the total energy of the system 
as a result of hole doping.
We do not consider the magnon excitation.
Thus the exchange terms in (\ref{hf}) above vanish 
due to the global SU(2) symmetry.

The diagonalization of (\ref{hammag}) yields 
the following generalized Harper's equation
which now incorporates electron correlations,
\beqn
\det(\bM-\ek\bI) = 0
\label{genhar}
\eeqn
with
\beqn
\bM = \bmat{c|c}
  \bT & \bU \\
  \hline
  \bU & -\bT
\emat~~,
\eeqn
where by setting $M_n = -2\cos(k_x+ng)$ and $t=1$,
\beqn
\bT = \bmat{ccccc}
  M_1     & -\emiky & 0      & 0       & -\eiky  \\
  -\eiky  & M_2     & \ddots & 0       & 0       \\
  0       & \ddots  & \ddots & \ddots  & 0       \\
  0       & 0       & \ddots & M_{q-1} & -\emiky \\
  -\emiky & 0       & 0      & -\eiky  & M_q    
\emat~~,
\eeqn
and
\beqn
\bU = \bmat{ccccc}
  -\sud & 0     & 0      & 0     & 0     \\
  0     & -\sud & 0      & 0     & 0     \\
  0     & 0     & \ddots & 0     & 0     \\
  0     & 0     & 0      & -\sud & 0     \\
  0     & 0     & 0      & 0     & -\sud 
\emat~~.
\label{U}
\eeqn
Here the matrix $\bT$ is associated with electron hopping,
involving phase modulation due to the external field,
and $\bU$, with the antiferromagnetic electron correlations.

\begin{figure}[bht]
\epsfxsize=6.0cm
\epsfysize=6.0cm
\centerline{\epsffile{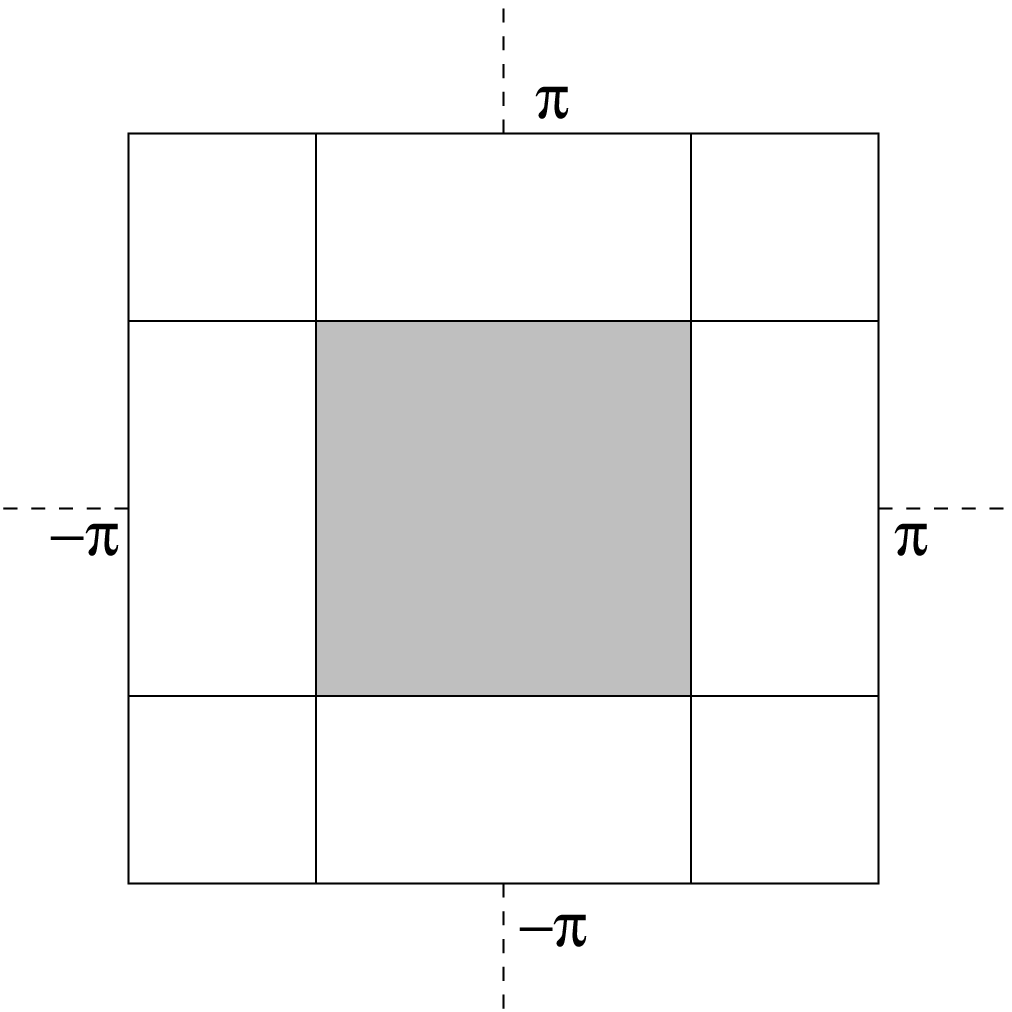}}
\centerline{(a)}

\vspace*{1cm}
\epsfxsize=6.0cm
\epsfysize=6.0cm
\centerline{\epsffile{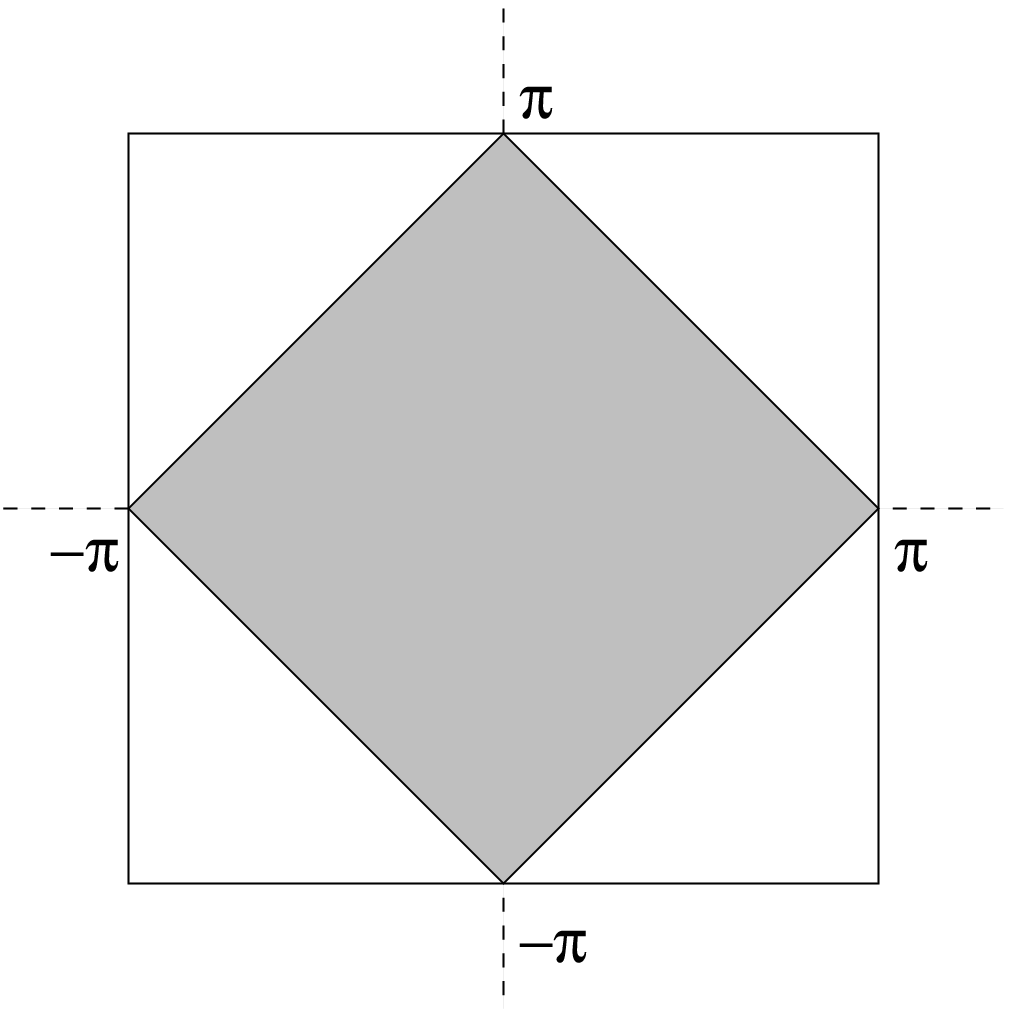}}
\centerline{(b)}
\caption{
  Reduced Brillouin zone (shaded region) 
  for the computation of energy dispersion relation with
  (a) half a flux quantum per plaquette, $\f{p}{q}=\f{1}{2}$,
  and
  (b) no magnetic field, $\f{p}{q}=0$.
  Here the elementary flux quantum $\phi_0$ is set to be a unity.
}
\label{brillouin}
\end{figure}

Now in order to realistically investigate 
the effect of an applied magnetic field 
on the two-dimensional system of antiferromagnetically correlated electrons
we consider a case of half a flux quantum per plaquette,
that is, $\f{\phi}{\phi_0}=\f{p}{q}=\f{1}{2}$.
The energy dispersion relation is then obtained as
\beqn
\Ekpm &=& \pm \sqrt{4t^2(\cos^2k_x+\cos^2k_y)+\left(\f{mU}{2}\right)^2}
  \nonumber \\
  && + \f{U}{2}(1-\d) - \mu~~. 
\eeqn
Here the staggered magnetization (AF order) $m$
is affected by the applied field.
The self-consistent mean field equations for $m$ and $\mu$ are given by
\beqn
1 &=& \int'
  \f{d^2k}{(2\pi)^2}\f{U}{\ek}
  \left[\tanh\left(\f{\Ekp}{2T}\right) 
  - \tanh\left(\f{\Ekm}{2T}\right)\right] \label{fluxhalfeq1}~~,\\
\d &=& 2\int'
  \f{d^2k}{(2\pi)^2}\left[\tanh\left(\f{\Ekp}{2T}\right)
  +\tanh\left(\f{\Ekm}{2T}\right)\right] \label{fluxhalfeq2}~~,
\eeqn
where the prime denotes integration over the reduced Brillouin zone of 
$(k_x,k_y)$ with $-\f{\pi}{2}\le k_x, k_y\le\f{\pi}{2}$,
as shown in Fig. \ref{brillouin}(a).

\begin{figure}[thb]
\epsfxsize=6.0cm
\epsfysize=6.0cm
\centerline{\epsffile{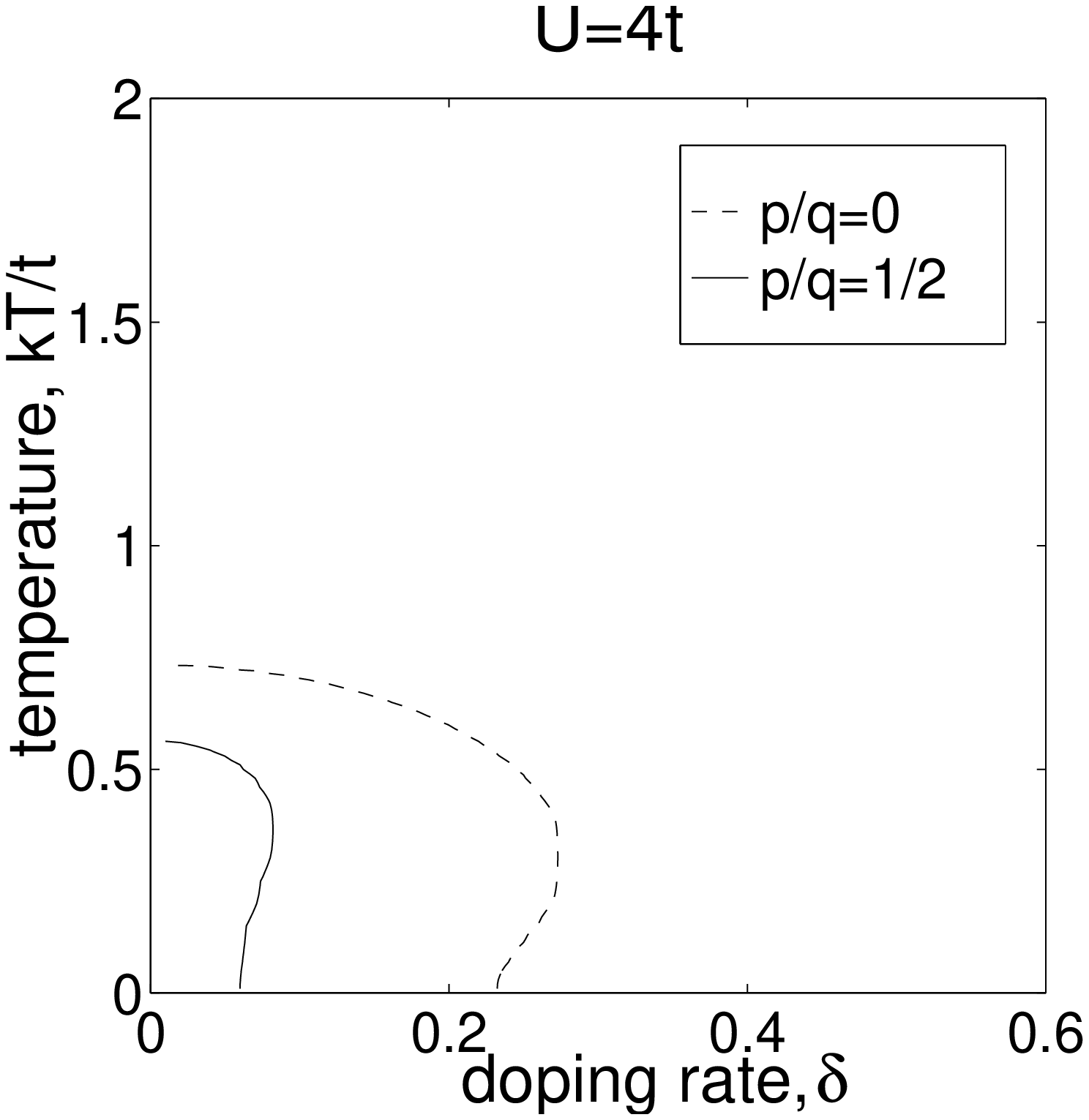}}
\centerline{(a)}

\epsfxsize=6.0cm
\epsfysize=6.0cm
\centerline{\epsffile{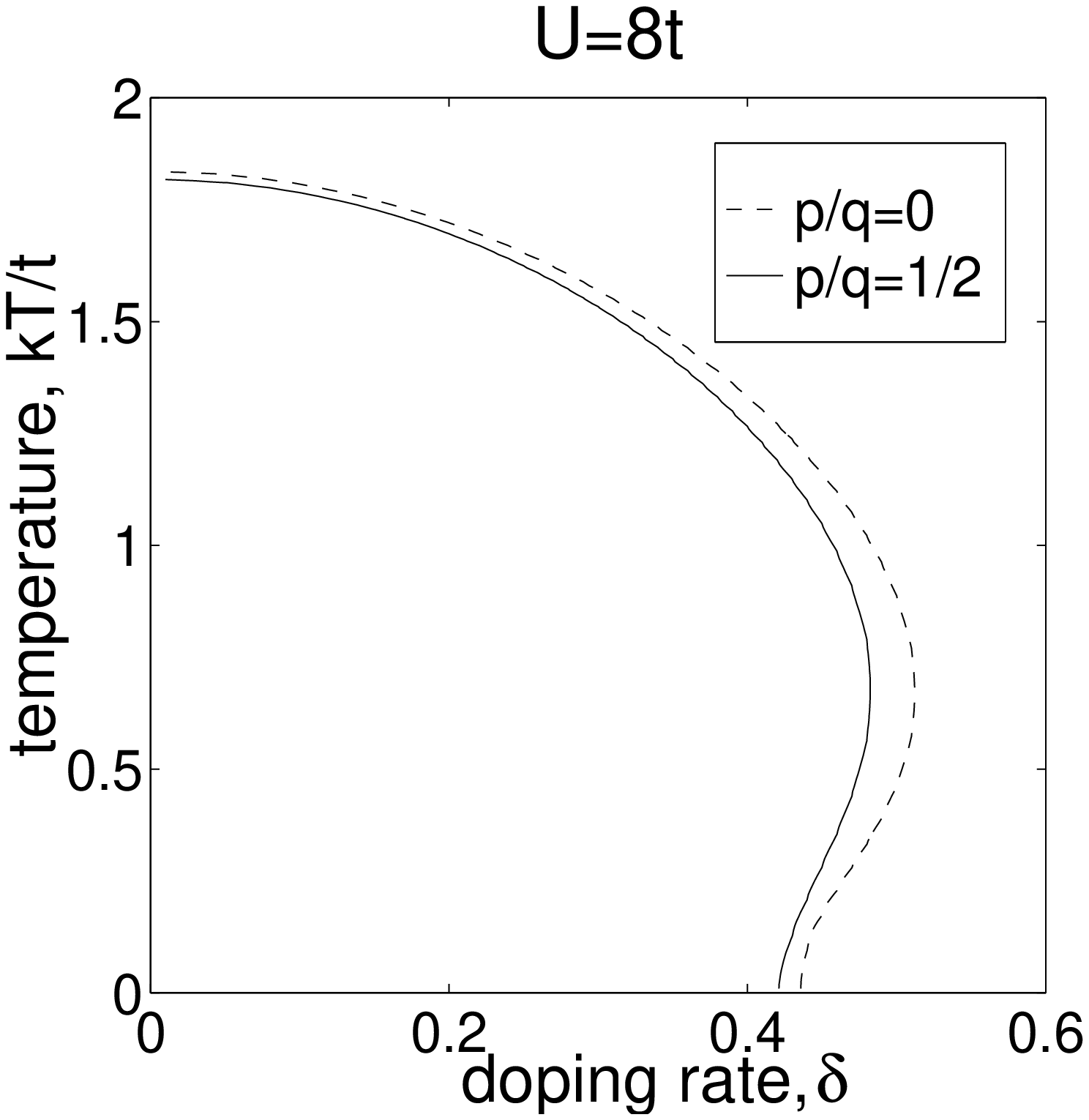}}
\centerline{(b)}
\caption{
  Phase diagram of the staggered magnetization (AF order)
  in the $(T,\d)$ plane with
  (a) $U=4t$ (b) $U=8t$.
  The dashed and solid lines indicate the boundary of antiferromagnetic states 
  for the magnetic flux quanta per plaquette,
  $\f{p}{q}=0$ and $\f{p}{q}=\f{1}{2}$
  respectively.
  Here the elementary flux quantum $\phi_0$ is set to be a unity.
}
\label{diagram}
\end{figure}
In order to examine the dependence of the external magnetic field 
(the solid lines in Fig. \ref{diagram})
on the staggered magnetization (AF order) $m$
in the $T$-$\d$ plane
we solved numerically
Eqs. (\ref{fluxhalfeq1}) and (\ref{fluxhalfeq2}).
We find that the reentrant behavior of the staggered magnetization
(AF order) persists at a certain region of doping rate $\d$
despite the application of the external magnetic field.
The reentrant behavior appears at smaller doping rates
for weaker correlations (see the case of $U=4t$),
as shown in Fig. \ref{diagram}.
The domain of antiferromagnetic phase is smaller
compared to the case of zero external field 
(the dotted lines in Fig. \ref{diagram}).
The band gap $mU$ owing to the antiferromagnetic correlations gets smaller 
in the presence of the external magnetic field
as a consequence of reduced magnetization $m$.
Such effect of reduction was predicted to be much larger with $U=4t$ 
than with $U=8t$, as shown in the figure.

\begin{figure}[tbh]
\epsfxsize=6.0cm
\epsfysize=6.0cm
\centerline{\epsffile{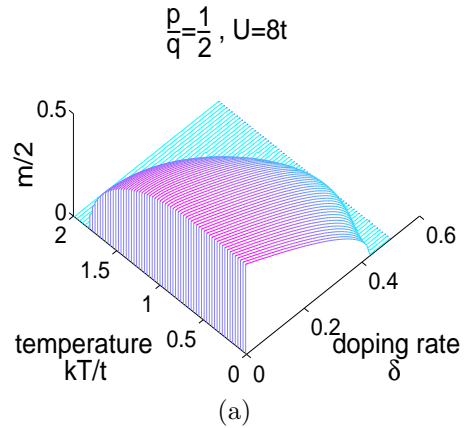}}
\centerline{(a)}

\epsfxsize=6.0cm
\epsfysize=6.0cm
\centerline{\epsffile{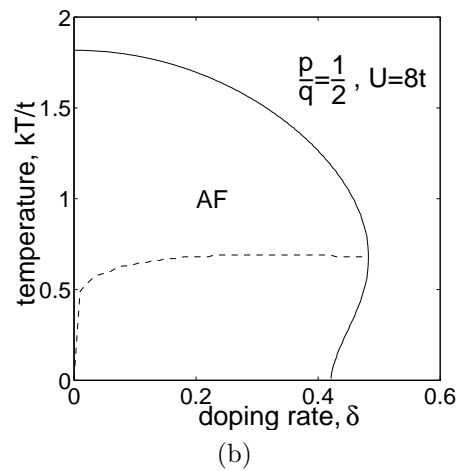}}
\centerline{(b)}

\epsfxsize=6.0cm
\epsfysize=6.0cm
\centerline{\epsffile{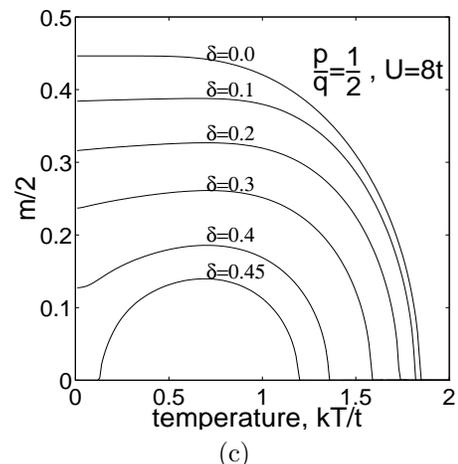}}
\centerline{(c)}
\caption{
  (a) Three dimensional plot of the staggered magnetization 
  as a function of temperature $T$ and doping rate $\d$
  for $U=8t$ in the absence of magnetic field.
  (b) The solid line represents a boundary 
  of the antiferromagnetic state.
  The dashed line denotes the temperature at which 
  the staggered magnetization $m$
  reaches the maximum value as a function of doping rate $\d$.
  (c) Temperature dependence of the staggered magnetization 
  (AF order) for various values of doping rate $\d$.
}
\label{reentrance}
\end{figure}

We now investigate the reentrant behavior of staggered magnetization
more in detail.
In Fig. \ref{reentrance}(a) we display the staggered magnetization
$m$ as a function of both temperature $T$ and doping rate $\d$.
At exactly half filling, i.e., $\d=0$, 
$m$ reaches the maximum at zero temperature.
We find that the temperature at which it
reaches a maximum value quickly increases 
and converges to a finite temperature near half filling,
as shown in Fig. \ref{reentrance}(b).
The reentrant behavior with $U=8t$ occurs more clearly 
at a region of higher doping rates
(say $0.42<\d<0.48$, as shown in Fig. \ref{reentrance}(b)).
We see more readily from Fig. \ref{reentrance}(c)
the variation of reentrant behavior with the increasing doping rate.
We find explicitly the rapid decrease of the AF order
(staggered magnetization)
with doping rate at a given temperature, e.g., $kT=1t$,
as shown in Fig. \ref{reentrance}(c).

Let us now consider the limiting case of 
vanishing  external magnetic field ($B=0$).
We obtain from (\ref{hammag}) the following expression,
\beqn
\lefteqn{
H = -2t\sumks (\cox+\coy)\cksd\cks
} \nonumber \\
&& -\f{mU}{2}\sumks \s c_{\bk+\bQ,\s}^\dr\cks
  +[\f{U}{2}(1-\d)-\mu]\sumks\cksd\cks 
\label{hamzero}
\eeqn
as a result of $\bg={\bf 0}$.
The dispersion relation is then,
by diagonalizing the Hamiltonian above \cite{schrieffer},
\beqn
\Ekpm &=& \pm \sqrt{4t^2(\cox + \coy)^2 + \left(\f{mU}{2}\right)^2}
  \nonumber \\
&& + \f{U}{2}(1-\d) - \m~~.
\eeqn
The self-consistent equations for $m$ and $\mu$ are, in form,
the same as (\ref{fluxhalfeq1}) and (\ref{fluxhalfeq2}) above.
However the integral is now over the reduced Brillouin zone of
$(k_x,k_y)$ with $|k_x|+|k_y|\le\pi$ 
as shown in Fig. \ref{brillouin}(b).

Away from half filling the N\'{e}el temperature is generally lower.
Now for the limiting case of the zero external field
(the dotted lines in Fig. \ref{diagram}),
we find that there exists a reentrant behavior
with a larger domain of staggered magnetization (AF order)
in the $T$-$\d$ plane
compared to the case
of the applied external field (the solid lines in Fig. \ref{diagram}).
In qualitative agreement with our present results,
Halvorsen et al. \cite{halvorsen} also showed the reentrant behavior
using a different approach.
They used the Hubbard model 
within the self-consistent second-order weak $U$-perturbation treatment,
thus allowing to study only a limited region of small $U$.
Lately, Inaba et al. \cite{inaba} found 
qualitatively a similar reentrant behavior
using the $t$-$J$ Hamiltonian in the slave-boson representation,
permitting to investigate only the large $U$ limit case.
On the other hand, 
our present approach allows to examine the full range of $U$,
although we are not able to display all details 
due to a limited space.
Further, unlike our present case,
their studies were limited only to 
the case of zero external magnetic field.

In the present study we derived the generalized Harper's equation 
which incorporates correlation effects between electrons.
Unlike other studies \cite{halvorsen,inaba}
we were able to examine 
as a function of both temperature and doping rate 
the variation of the staggered magnetization $m$
for the system of antiferromagnetically correlated electrons
under the external magnetic field.
By deducing the phase diagram in the $T$-$\d$ plane,
we explored staggered magnetization $m$
as a function of both temperature and doping rate
and found the reentrant behavior
even in the presence of the applied magnetic field.
Finally, it is of note that the staggered magnetization or the AF order 
for other values of magnetic flux 
$\frac{p}{q}\phi_0$
can be investigated in a similar manner.
Further studies regarding this problem will be resumed on.

One (S.H.S.S) of the authors is grateful to the SRC program (1996) of
the Center for Molecular Science at KAIST.
He acknowledges the financial supports of
the Korean Ministry of Education BSRI (1996)
and POSTECH/BSRI special programs.
He is also grateful to Professor Han-Yong Choi for assistance.

\end{document}